\documentclass[a4paper,11pt]{article}
\usepackage{pos}

\usepackage{amsmath,slashed,mathbbol,mathtools}

\newcommand{\rme}{\mathrm{e}}
\newcommand{\rmd}{\mathrm{d}}
\newcommand{\rmi}{\mathrm{i}}

\usepackage{xcolor}

\title{Quantum M2-branes and Holography}
\author*[a]{Jesse van Muiden}

\affiliation[a]{Abdus Salam Centre for Theoretical Physics, Imperial College London,\\
  Prince Consort Road, London SW7 2AZ, UK}

\emailAdd{jvanmuid@ic.ac.uk}

\abstract{We discuss the semi-classical quantisation of supersymmetric membranes in holographic geometries with asymptotic AdS$_4$ and AdS$_7$ boundary conditions. In AdS$_4$ geometries this quantisation prompts the need of ensemble changes when comparing bulk and boundary observables arising from such membranes. We also discuss how supersymmetric membranes localise to loci where the background Killing spinor turns chiral, circumventing the need of evaluating their zero-mode moduli integrals. Finally, we discuss a bulk analysis of an infinite tower of membrane instantons, or giant gravitons, in AdS$_7$ geometries, whose worldvolume dynamics effectively reduces to a quantum mechanical system. This allows us to test, in a holographic setting, the emergence proposal that perturbative supergravity data may be extracted from towers of membrane instantons.
}

\FullConference{Proceedings of the Corfu Summer Institute 2025 "School and Workshops on Elementary Particle Physics and Gravity" (CORFU2025)\\
27 April - 28 September, 2025\\
Corfu, Greece\\}

\begin{document}
\maketitle
\section{Introduction}
String theory has provided a precise setup to study holographic dualities through the AdS/CFT correspondence and its extensions. Testing these precise dualities provides the opportunity to learn about the non-perturbative nature of gauge theories and quantum gravity. Such tests can be done most rigorously, in generic regimes of parameter space, for supersymmetric observables computable through supersymmetric localisation in the dual quantum field theory. Arguably the simplest such observable is the partition function, for which the holographic duality tells us that
\begin{equation}\label{Eq: string partition function in holography}
	\log Z_{\text{QFT}} = \mathcal Z_\text{string}\,,
\end{equation}
where the logarithm is included to ensure we are comparing the generating functionals of connected diagrams in both theories. This QFT partition function can be refined through the insertion of extended operators, which in the gravity description involves extended strings and branes in the background geometry. In the setups we discuss the left-hand side  of this equation will already have been computed using supersymmetric localisation. The right hand side is still a notoriously complicated object, in particular since generic top-down holographic geometries involve non-trivial R-R and NS-NS fluxes. Formally at least, the sigma model approach to string theory does provide an approach to compute the string partition function
\begin{equation}\label{Eq: string partition function}
	\mathcal Z_{\text{string}}[g_{MN},\Phi,\ldots] = \sum_\chi \frac{1}{g_s^{\chi}} \int\mathcal D[h,\phi,\psi]\rme^{-S[h,\phi,\psi;g_{MN},\Phi,\ldots]}\,,
\end{equation}
where $\chi$ is the worldsheet Euler characteristic, and the target space metric, dilaton and fluxes function as sources in the string partition function. As a simplification the path integral can be expanded in saddle points at long wavelengths, where the leading saddle corresponds to a degenerate genus zero string which upon semi-classical quantisation leads to the supergravity action, and is expected to also determine its higher derivative corrections \cite{Fradkin:1984pq,Fradkin:1985fq}\footnote{This sigma-model approach has been applied in the bosonic string to compute the $R^2$ correction from two loops \cite{Andreev:1990iv}.} 
\begin{equation}\label{Eq: Saddle expansion of string partition function}
\begin{aligned}
	\mathcal Z_{\text{string}} =&\, \sum_{\text{saddles}} \rme^{-S_{\text{cl}}}(Z_{\text{1-loop}} + \ldots)
	=\, -S_{\text{sugra}} + S_{\text{HD}} + \sum_\text{inst.} \rme^{-S_{\text{cl}}}(Z_{\text{1-loop}} + \ldots) + \mathcal O(g_s^0)\,,
\end{aligned}
\end{equation}
where the remaining sum is over worldsheet instantons and we have suppressed higher genus corrections. Note that deriving the supergravity action was much subtler than presented here. It arises only after removing a residual Mobius group gauge symmetry and regulating logarithmic divergences whose coefficient actually encodes the Lagrangian density of the supergravity theory \cite{Tseytlin:1987ww,Tseytlin:1988tv}. The action subsequently arises once the Lagrangian density is integrated over the zero mode moduli of the string. This latter point is of importance to us as it depicts how one should interpret the localisation of the supergravity action within string theory.

In what follows our main interest will be the non-perturbative instanton contributions to the string partition function. This has lead to some interesting results in four independent contexts, which we will now quickly review and discuss in more detail in the sections below.

\paragraph{I. Equivariant localisation.} Considerable effort has gone into exploring the applicability of equivariant localisation in supergravity theories. Starting with \cite{BenettiGenolini:2019jdz} where the authors noted that the on-shell action in four-dimensional gauged supergravity can be evaluated as a sum over a collection of fixed points and surfaces. This was later formalised to be a direct consequence of the Atiyah-Berline-Bott-Vergne fixed point formula \cite{BenettiGenolini:2023kxp}, which was subsequently applied to a wide range of supergravity theories and their observables. The associated fixed nuts and bolts are determined by a background Killing spinor which turns chiral at those loci. In fact it was noted in \cite{Genolini:2021urf} that also the higher derivative corrections to supergravity can be evaluated on these fixed points. In the context of string theory, and the saddle point approximation of its partition function discussed above, the ABBV formula acts directly on the integral over zero-mode moduli of the string. Additional support for this fact, and that indeed the entire string partition function equivariantly localises on supersymmetric holographic backgrounds, was provided in \cite{Gautason:2025per} where it was shown that also the integral over zero-mode moduli of subleading saddles in the string partition function localises to loci where the background Killing spinor turns chiral.

\paragraph{II. A sum over membrane saddles.} In \cite{Gautason:2023igo} we have taken the saddle point approximation to the string partition function \eqref{Eq: Saddle expansion of string partition function} and used it to test holographic dualities at non-perturbative levels. In particular, it was shown that non-perturbative contributions to the sphere partition functions of the ABJM theory (at large 't Hooft coupling) and five-dimensional super-Yang-Mills, i.e. the left hand side of \eqref{Eq: string partition function in holography}, can be holographically matched precisely to quantised worldsheet instantons in the respective dual geometries. These results were subsequently used in two interesting papers \cite{Beccaria:2023ujc,Beccaria:2023sph}, which uplifted the quantised worldsheet instanton to membrane instantons in M-theory and showed that their semi-classical quantisation correctly matched non-perturbative contributions to the ABJM partition function at finite $k$ and the (2,0) partition function on $S^5\times S^1$.\footnote{These are the appropriate strong coupling limits of the ABJM and 5d SYM theories with M-theory duals.} We in turn use these results as an inspiration, in combination with the saddle point approximation in \eqref{Eq: Saddle expansion of string partition function}, to study the existence of a putative Euclidean membrane path integral, evaluated as a sum over saddles
\begin{equation}\label{Eq: membrane path integral}
	\mathcal Z_{\text{M2}}[g_{MN},A_3] = \sum_{\text{saddles}} \rme^{-S_{\text{cl}}}(Z_{\text{1-loop}} + \ldots)
	=\,\color{red} -S_{11d} + \ldots\color{black} + \color{violet} \sum_\text{inst.} \rme^{-S_{\text{cl}}}(Z_{\text{1-loop}} + \ldots)\color{black} \,,
\end{equation}
which is once again a functional of the background sources $g_{MN}$ and $A_3$. Contemplating the existence of such a membrane path integral as a sum over saddles is of course not new, we believe however that holography provides important new opportunities to study its validity. When doing so one should keep in mind that there are a number of obstructions, which are best worded in the divertimento of \cite{Harvey:1999as}. First, to this day there are no methods directly quantising the membrane in any analog to \eqref{Eq: string partition function}. The subsequent obstruction is that there is no consistent procedure to semi-classically quantise a degenerate membrane or how to derive the low-energy graviton dynamics, highlighted in red. Thus finding the eleven-dimensional supergravity action as the leading saddle contribution, is out of reach, let alone its higher derivative corrections. Our focus instead lies with the infinite sum over subleading membrane saddles, highlighted in violet. In the context of holography the quantisation of these membrane saddles can be explicitly verified to be consistent through a comparison to supersymmetrically localised field theory observables. This allows one to test the holographic equality
\begin{equation}\label{Eq: membrane partition function in holography}
	\log Z_{\text{QFT}} = \mathcal Z_\text{M2}[g_{MN},A_3]\,,
\end{equation}
and provides evidence that a membrane approach to M-theory as in \eqref{Eq: membrane path integral} might prove fruitful even with all the caveats in mind.
\paragraph{III. Ensembles.} Let us now take a step back and comment on a surprising feature when one takes a proposal like \eqref{Eq: membrane path integral} and \eqref{Eq: membrane partition function in holography} seriously. Namely, the quantisation of supersymmetric membranes is done in the presence of background sources, which are the configurations of the background supergravity fields to which the membrane couples, in particular the metric $g_{MN}$ and the three-form gauge potential $A_3$. In the context of holography one thus has to ensure that the field theory observable is computed with the same sources fixed. To emphasise the importance of this fact we can take the well-studied example of AdS$_4 \times S^7/\mathbf{Z}_k$ holography, although the discussion is much more general than this \cite{Gautason:2025plx}. The dual field theory in this setup is the ABJM theory, consisting of two gauge nodes of rank $N$ with opposite CS-levels and four bi-fundamental chiral multiplets with a quartic superpotential \cite{Aharony:2008ug}. The CS-level is to be directly identified in the bulk with the rank of the $\mathbf{Z}_k$ orbifold. The rank of the gauge group is as usual identified with the quantised flux number in the bulk
\begin{equation}
	N = \frac{1}{(2\pi \ell_p)^6} \int_{S^7/\mathbf{Z}_k} G_7\,.
\end{equation}
The holographic equality of generating functionals thus imposes the bulk path integral to be computed at fixed value of the six-form gauge potential $A_6$. This is however not the same ensemble in which the membrane is quantised, and thus bulk observables arising from quantum membranes such as membrane instantons and in fact the supergravity action, when taking \eqref{Eq: membrane path integral} and \eqref{Eq: membrane partition function in holography} seriously, require an ensemble change when compared to the corresponding field theory observables.\footnote{Note that eleven-dimensional supergravity cannot be written in a local form purely in terms of the gauge potential $A_6$.} It was discussed in \cite{Gautason:2025plx} that the change in ensemble is performed in a standard way for two canonically conjugate variables, which in this case are the $A_6$ and $A_3$ potentials which function as the particle number and chemical potential respectively. In the simplest AdS$_4$ setup, with a round $S^3$ boundary, we identify the canonical conjugate variables with $A_3 \sim \mu$ and $A_6 \sim N$, and relate the partition functions in the two ensembles by\footnote{For different boundary conditions the relation between the ensembles can get more involved depending on the non-trivial cycles wrapped by the respective gauge potentials.}
\begin{equation}
	\rme^{\mathcal Z_{\text{M2}}[\mu,k]} = \sum_{N=0}^\infty Z_{\text{ABJM}}[N,k]\rme^{\mu N}\,.
\end{equation}
Indeed, we discuss non-trivial evidence that observables computed from membranes and supergravity need correspond to observables in a grand canonical ensemble on the boundary.

Note that this subtlety of ensemble choices in holography determined by fixed fluxes has also been discussed in the context of AdS$_3$ holography, see e.g. \cite{Eberhardt:2020bgq,Aharony:2024fid}. In general these subtleties are related to the application of hodge dualities in a partition function, forcing a sum over flux sectors, as recently reviewed in the context of four-dimensional GR with an axion \cite{Witten:2026twr}.

\paragraph{IV. Emergence.} Finally, let us come back to the perturbative supergravity contributions in the membrane partition function. As mentioned, there is currently no technology available to understand if these can or should be determined from a semi-classical quantisation procedure of a degenerate membrane, as was the case for the string, i.e. \eqref{Eq: Saddle expansion of string partition function}. Attempts have been made to circumvent such a quantization altogether by instead extracting the supergravity dynamics from membrane instantons, through the so-called emergence proposal, see \cite{Hattab:2024ssg,Blumenhagen:2024lmo} and references therein. These approaches have shown some merit in the context of topological string theory on Calabi-Yau three-folds, which in turn can be uplifted to eleven dimensions. In this context it has long been understood how integrating out world-sheet instantons on the CY manifold computes F-term corrections in the lower-dimensional gravitational theories \cite{Candelas:1990rm,Gopakumar:1998ii,Gopakumar:1998jq}. The emergence proposal states that also the leading supergravity data can be computed in this method. If this approach holds it implies that from an analytic control of finite-size membrane instantons one is able to extract information of the degenerate membrane, circumventing the obstacle of its quantisation. We will briefly discuss the viability of this method and its application in holographic geometries.
\paragraph{Comment.} This manuscript is prepared for the proceeding of the Corfu 2025 Workshop on Quantum Gravity and Strings and is based on published works \cite{Gautason:2023igo,Gautason:2025per,Gautason:2025plx} and upcoming work \cite{futurepaper,futurepaperII}.
\section{Quantising the Membrane}
Before moving on to discuss explicit holographic successes concerning the semi-classical quantisation of membranes we first summarize the methods we use in a more general framework. We will work with membranes in eleven dimensions, whose consistent geometries are determined by eleven-dimensional supergravity
\begin{equation}\label{Eq: 11d action}
	S_{11d} = \frac{2\pi}{(2\pi \ell_p)^9} \int \star (R + |G_4|^2) + \frac16 A_3 \wedge G_4 \wedge G_4 \,.
\end{equation}
Concerning the fermions, we only need the supersymmetry variations 
\begin{equation}
	\delta \Psi_M = \Big(\nabla_M + \frac{1}{24}\Gamma_M\slashed{G}_4- \frac{1}{8}\slashed{G}_4\Gamma_M\Big)\epsilon = 0\,,
\end{equation}
where $\epsilon$ is the background Killing spinor. The coupling to supersymmetric membranes of charge $q_{\text{M2}}$ arises through the BST action \cite{Bergshoeff:1987cm} 
\begin{equation}
	S_{\text{M2}} = \frac{2\pi}{(2\pi \ell_p)^3} \left(\int \rmd ^3 \xi \sqrt{\hat g} +\rmi q_{\text{M2}}\int \hat A_3 + \int \rmd ^3 \xi \sqrt{\hat g} \,\bar \Theta ( \hat{\slashed{\nabla}} + \frac18 (\slashed{G}_4 - \hat\Gamma^\alpha \slashed{G}_4\hat \Gamma_{\alpha}))\Theta\right)\,,
\end{equation}
where we have explicitly introduced hats to denote the pulled back objects from the target space, although from now on we will omit them for notational simplicity. We will work in static gauge and have already fixed the $\kappa$-gauge to $\Gamma_{\kappa} \Theta =\rmi  \Theta$, where $\Gamma_\kappa$ is the rank three $\Gamma$-tensor along the worldvolume $\Gamma_{\kappa} = \frac{1}{3!} \varepsilon^{\alpha\beta\gamma}\Gamma_{\alpha\beta\gamma}$. The same tensor dictates the compatibility of the membrane supersymmetry with that of the target space through
\begin{equation}\label{Eq: membrane kappa symmetry}
	(1-\rmi q_{\text{M2}}\Gamma_{\kappa}) \epsilon = 0\,.
\end{equation}
To semi-classically quantise these membranes we compute the fluctuations on its worldvolume, and at one loop the quadratic fluctuations are sufficient. To compute these fluctuations we expand the worldvolume action in Riemann normal coordinates to find
\begin{equation}\label{Eq: general quadratic action M2 brane}
	S_{\text{M2}}^{(2)} = \int \rmd^3 x \sqrt{{{g}}} \frac12 (\mathcal D^\alpha \phi \mathcal D_\alpha \phi -  \phi\, m\, \phi + \bar \theta \Gamma^\alpha {\mathcal D_\alpha} \theta)\,,
\end{equation}
where the covariant derivative is defined to include a term coming from the WZ action 
\begin{equation}
\begin{aligned}
	\mathcal D_\alpha^{ij} =&\, \delta^{ij}\nabla_\alpha + \omega_\alpha^{\phantom{\alpha}ij} + \frac14 \varepsilon_{\alpha\beta\gamma} G^{\beta\gamma ij}\,,\\
	m_{ij} =&\, R^\alpha_{\phantom{\alpha}i\alpha j} + K_{\alpha\beta i} K^{\alpha\beta}_j - \frac{\rmi}{3!} \varepsilon^{\alpha \beta \gamma} \nabla_i G_{j \alpha\beta\gamma} - \frac18 G_{\alpha\beta ik}G^{\phantom{j}\alpha\beta k}_{j}\,.	
\end{aligned}
\end{equation}
In the examples that follow we diagonalise these fluctuations on the worldvolume into four complex scalars and four fermions, whose respective kinetic operators take the form
\begin{equation}
	\mathcal K = -D^2 + M^2\,,\qquad \mathcal D = \rmi \slashed D + M\,, \quad \text{with} \quad D = \nabla -\rmi  q_i\mathcal A^i\,,
\end{equation}
in terms of which their one-loop partition functions take the standard form
\begin{equation}
	Z_{\text{1-loop}} = Z_{\text{0-modes}} \frac{\prod_f \sqrt{\text{det}'\mathcal D} }{\prod_b \sqrt{\text{det}'\mathcal K} }\,,
\end{equation}
which has to be appropriately regulated. The primes denote that we collected all possible zero-modes into $Z_{\text{0-modes}}$.
\section{Membranes in AdS$_4 \times S^7/\mathbf{Z}_k$}
Our first case study of quantum membranes will be in the case of asymptotically locally AdS$_4 \times S^7/\mathbf{Z}_k$ geometries. Depending on the boundary conditions, these geometries describe different observables in the ABJM theory. All of the solutions can be uplifted from minimal gauged $\mathcal N=2$ supergravity in four dimensions, whose bosonic action equals
\begin{equation}\label{4daction}
	S = -\frac{1}{16\pi G_{N}^{(4)}} \int \star_4 \bigg(R_4 + 6 - \frac14 F_{\mu\nu}F^{\mu\nu}\bigg)\,,
\end{equation}
where $F = \rmd A$ and the gravitino BPS equation equals
\begin{equation}
	\delta_\eta \psi_\mu  =  \Big(\nabla_\mu - \rmi \frac{A_\mu}{2} + \frac12 \gamma_\mu   + \frac{\rmi}{4} \slashed{F} \gamma_\mu\Big)\eta = 0\,,
\end{equation}
with $\eta$ the background Killing spinor. Indeed, solutions to this theory can be uplifted to solutions of eleven-dimensional supergravity compactified on $S^7$ \cite{Gauntlett:2007ma} with a metric and flux\footnote{An important remark is that the IIA limit is taken by acing with $\mathbf{Z}_k$ inside $\mathbf{CP}^3$. An explicit parametrisation of the associated frames can be found in \cite{Gautason:2025per}, where also the $\mathbf{Z}_k$ action is spelled out.}
\begin{equation}\label{Eq: 11d metric uplift}
	\rmd s_{11}^2 = L^2 (\rmd s_4^2 + 4 \rmd s_{\textbf{CP}^3}^2 + (\rmd y + 2\sigma + A/2)^2)\,,
	\quad G_4 =\rmi\, L^3 (3 \text{vol}_4 + 2 J \wedge \star_4 F)\,,
\end{equation}
where $2J = \rmd \sigma$ is the K\"ahler form on $\mathbf{CP}^3$. Importantly, also the BPS equation can be consistently uplifted to eleven dimensions upon taking a factorised ansatz for the background Killing spinor:
\begin{equation}\label{Eq: 11d spinor}
	\epsilon = \eta \otimes \chi^+ + \eta^c \otimes \chi^-\,, \quad \Rightarrow \quad \delta_{\epsilon} \Psi_\mu = \delta_{\eta} \psi_\mu \otimes \chi^+ + \delta_{\eta^c} \bar \psi_\mu \otimes \chi^-\,,
\end{equation}
where $\chi^{\pm}$ are Killing spinors on $S^7/\mathbf{Z}_k$
\begin{equation}
	\chi^{\pm} =  \rme^{\rmi y \gamma_{11}}\frac{1 \pm \gamma_{11}}{2}\,  \frac{1 - \rmi \gamma_{\textbf{P}} \gamma_{11}}{2} \,\frac{1 - \rmi \gamma_{\tilde{\textbf{P}}}\gamma_{11}}{2} \chi_0\quad \Rightarrow\quad  \slashed{J} \chi^{\pm} \propto \pm \chi^{\pm}\,,
\end{equation}
with $\gamma_{\textbf{P}}$ and $\gamma_{\tilde{\textbf{P}}}$ rank two gamma matrices along the two-cycles $\textbf{P}^1 \times \tilde{\textbf{P}}^1 \subset \textbf{P}^3$. The explicit construction of these spinors can be found in \cite{Gautason:2025per}.
\paragraph{Membrane instantons.} Membranes wrapping three-cycles inside $S^7/\mathbf{Z}_k$ are subleading saddles in \eqref{Eq: membrane path integral} contributing non-perturbatively in $L/\ell_p$ to the path integral. To ensure supersymmetry we solve \eqref{Eq: membrane kappa symmetry} which, with the Killing spinor in \eqref{Eq: 11d spinor}, reduces to the following chirality constraint 
\begin{equation}\label{Eq: chirality constraint}
	\rmi \gamma_{\text{M2}} \chi^{\pm} = \pm\chi^{\pm} \,,\quad  \gamma_{(4)} \eta = \pm q_{\text{M2}} \eta\,, 
\end{equation}
where $\gamma_{\text{M2}}$ is the seven-dimensional gamma matrix pulled back onto the worldvolume and $\gamma_{(4)}$ is the four-dimensional chirality operator. This constraint reduces the zero-mode integral of the quantised membranes to a sum over fixed points and surfaces where the four-dimensional spinor turns chiral. This is entirely consistent with the equivariant localisation results in four-dimensional gauged supergravity \cite{BenettiGenolini:2023kxp} and suggests that in fact the entire M-theory partition function localises.

Inside $S^7/\mathbf Z_k$ there are two independent three-cycles which solve the $\kappa$-projection equation \cite{Gautason:2025per}. Their worldvolume metrics and on-shell actions take a universal form, namely 
\begin{equation}\label{Eq: metric and on shell action of membrane in AdS4}
	\rmd s_{\text{M2}}^2 =  L^2 \rmd s^2_{S^3/\mathbf{Z}_k} = L^2(\rmd \theta^2 + \sin^2\theta + (\rmd \varphi - \cos\theta \rmd \phi)^2)
	\,,\qquad S_{\text{M2}} = \frac{4 L^3}{k \ell_p^3 }\,.
\end{equation}
\subsection{Empty AdS$_4$}
Let us first discuss the simplest case where we take $A=0$ in the four-dimensional gauged supergravity theory such that the eleven dimensional geometry is simply AdS$_4 \times S^7/\mathbf{Z}_k$ with an $S^3$ boundary. The advantage of this simple solution is that the partition function of the dual gauge theory has been computed non-perturbatively as a function of $N$ and $k$ in a series of papers \cite{Drukker:2010nc,Marino:2011eh,Drukker:2011zy,Hatsuda:2012dt,Hatsuda:2012hm,Hatsuda:2013gj}. In a large $N$ expansion its partition function was shown to resum to an Airy function up to exponentially suppressed corrections
\begin{equation}
	Z^{S^3}_{\text{ABJM}}[N,k] = \mathcal C_k^{-1/3} \rme^{\mathcal A_k} \text{Ai}[\mathcal C_k^{-1/3}(N-\mathcal B_k)] + \sum_{m,n} f_{m,n}(N,k) \rme^{-2\pi(n\sqrt{N/k} + m \sqrt{N k})}\,,
\end{equation}
where 
\begin{equation}
	\mathcal C_k = \frac{2}{\pi^2 k}\,,\quad \mathcal B_k = \frac{1}{3k} + \frac{k}{24}\,,\quad \mathcal A_k = \frac{2\zeta(3)}{\pi^2 k }(1 - \frac{k^3}{16}) + \frac{k^2}{\pi^2} \int\limits_0^\infty\rmd x \frac{x \log (1 - \rme^{-2x})}{\rme^{k x} - 1}\,,
\end{equation}
and the non-perturbative contributions, controlled by the coefficients $f_{m,n}(N,k)$, were argued to also take the form of Airy functions \cite{Hatsuda:2012dt}. These instantonic contributions can be distinguished into three different classes which owe their names to their respective string theory interpretations \cite{Drukker:2011zy,Hatsuda:2013gj}: worldsheet instantons for $(n,0)$, D2-brane instantons for $(0,m)$, and boundstate instantons for $(n,m)\neq 0$. In what follows we will only be interested in worldsheet instantons which dominate as long as their instanton level does not exceed the orbifold number $k$.\footnote{Note that in fact these instantons arise from multicoverings inside $S^7/\mathbf{Z}_k$, which can only be regular if the degree of their covering map does not exceed the orbifold number.} 

Due to the high amount of control, the sphere partition function of the ABJM theory is an ideal candidate to holographically test the existence of a membrane path integral as a sum over saddles as in \eqref{Eq: membrane path integral}, and its different consequences as discussed in the introduction. Let us start with verifying the holographic duality at the level of the eleven-dimensional two-derivative supergravity action, which is naively expected to equate to the large $N$ field theory free energy.\footnote{We take the field theory free energy to equal $F = -\log Z$, and we will from now on omit the super- and subscripts on the $F$ for simplicity of presentation.}  The latter equals
\begin{equation}
	\lim_{N\rightarrow \infty} F[N,k] = \frac{(2 k)^{1/2} \pi}{3}N^{3/2}\,.
\end{equation}
To compare to the gravitational on-shell action one has to change from field theory to gravitational variables, canonically done through a bulk flux quantisation, which in our setup gives
\begin{equation}\label{Eq: flux quantisation}
	N = \frac{1}{(2\pi \ell_p)^6} \int_{S^7/\mathbf{Z}_k} G_7  = \frac{2}{\pi^2 k} \frac{L^6}{\ell_p^6} \quad \Rightarrow \quad \lim_{N\rightarrow \infty} F[N,k] = \frac{4}{3 \pi^2 k}\frac{L^9}{\ell_p^9}\,.
\end{equation}
Surprisingly, and importantly, this answer does not equal the eleven-dimensional on-shell supergravity action, which instead takes the value
\begin{equation}
	-S_{11d} = \frac{2}{3\pi^2 k} \frac{L^9}{\ell_p^9}  \,,
\end{equation}
and is off by a factor $2$ from the field theory answer. The resolution to this holographic mismatch was argued in \cite{Gautason:2025plx} to lie with the fact that the membrane partition function instead is computed at fixed $A_3$, and not fixed $A_6$ corresponding to fixed $N$. And thus the supergravity action, which in fact is computed at fixed $A_3$, is to be compared to the field theory only upon an ensemble change. To change ensembles at the level of the partition function we have to sum over flux sectors
\begin{equation}\label{Eq: inverse laplace of free energy}
	\rme^{J(\mu,k)} = \sum_{N=0}^\infty \rme^{\mu N} \rme^{-F[N,k]}\,,\quad \leftrightarrow \quad \rme^{-F[N,k]} = \frac{1}{2\pi i}\int_{\mathcal{C}} d\mu \, e^{J(\mu,k)-\mu N}\,,
\end{equation}
where $J(\mu,k)$ is the grand canonical potential. We have presented the change of ensembles here in the standard fashion by going between fixed particle number and fixed chemical potential. The gravitational ensembles correspond to either fixing the gauge potential $A_6$ or $A_3$, respectively, and their relation to the field theory parameters can be directly understood in the dominant supergravity saddle through the flux kinetic terms
\begin{equation}
	\frac{2\pi}{(2\pi \ell_p)^9}\int G_4\wedge \star G_4 = \mu N \quad \text{where} \quad \mu = \frac{L^3}{\ell_p^3}\,.
\end{equation}
Note that this is indeed consistent with the standard expression for the expectation value of the particle number at fixed chemical potential:
\begin{equation}
	\langle N \rangle = \frac{\partial J(\mu,k)}{\partial \mu}\,,
\end{equation}
and in the saddle point approximation we find that $\mu$ and $N$ are related by $\mu_\star^2 = \frac{N k \pi^2}{2}$.

The simplicity of the transform going between canonical and grand canonical ensembles in this setup is directly related to the simplicity of the bulk geometry where the flux sectors are determined by a single cycle on which $G_7$ and thus $N$ is defined. In more general setups, which we will discuss below, the procedure of changing ensembles can be more involved depending on the different non-trivial flux sectors of $A_3$ and $A_6$.

Leaving these subtleties aside for now we note that for the ABJM theory on $S^3$ the grand canonical potential was in fact already computed in the early days of precision holography, where it was found that \cite{Fuji:2011km}
\begin{equation}\label{Eq: Grand potential}
	J(\mu,k) = \frac{\mathcal C_k}{3}\mu^3 + \mathcal B_k \mu + \mathcal A_k + \sum_{n,m} \tilde f_{m,n}(\mu,k)\rme^{-4(m\mu/k - n \mu/2)}\,,
\end{equation}
where again the coefficients $\tilde f_{m,n}(\mu,k)$ can be split in three distinct classes depending on their string theory interpretations. The point that we want to convey here is that the associated chemical potential $\mu$ has a direct gravitational meaning as the fixed value of the background field $A_3$ to which the supersymmetric membranes couple. And indeed, this directly resolves the holographic mismatch we saw before since
\begin{equation}\label{Eq: match between sugra and grand canonical potential}
	-S_{11d} = \frac{2\mu^3}{3\pi^2 k} = \lim_{\mu\rightarrow \infty} J(\mu,k)\,,
\end{equation}
which we view as a first piece of non-trivial evidence verifying that bulk observables coming from quantised membranes correspond in the boundary to observables in a grand canonical ensemble.

Away from the supergravity saddle point we propose that in fact the entire membrane partition function is to be compared holographically to the grand canonical potential
\begin{equation}\label{Eq: Z = J}
	\mathcal Z_{\text{M2}}[g_{MN},A_3] = J(\mu,k)\,,
\end{equation}
and not the free energy. To provide further evidence for this equivalence it seems imperative to compute the eight derivative corrections to the supergravity action, which based on dimensional analysis are expected to fully determine the coefficient $\mathcal B_k$ \cite{Gautason:2025plx}. The issue is that these corrections are not fully known, making this a particularly difficult task. Instead, we will study the membrane instantons and their corresponding partition functions to further solidify the equality in \eqref{Eq: Z = J}. Before doing so let us note some important consequences in supergravity when taking this holographic equivalence seriously. The $\mu^3$ and $\mu^1$ terms in \eqref{Eq: Grand potential} correspond to the on-shell supergravity action and its eight-derivative correction respectively, and based on \eqref{Eq: Grand potential} we conclude that there are no further local supergravity corrections to this answer. Schematically, we have that
\begin{equation}\label{Eq: Sugra corrections Airy conjecture}
	 \int 
	\sqrt{g}(R +\ldots + R^4 +\ldots) \sim \frac{\mathcal  C_{k}}{3} \mu^3 + \mathcal B_k \mu\,,
\end{equation}
where the ellipses denote supersymmetric completions of the relevant curvature terms. There is one more perturbative contribution, the $\mu^0$ term $\mathcal A_k$. Based on dimensional analysis one can rule out its origin as a local supergravity correction. Instead, such constant contributions are expected to arise from one-loop corrections in M-theory. In fact, there is a close analogy with topological string theory where this contribution is known as the so-called constant map describing a trivial embedding of a membrane inside the Calabi-Yau \cite{Gopakumar:1998ii,Gopakumar:1998jq,Dedushenko:2014nya}.\footnote{In fact this analogy is the origin of its name in the ABJM sphere partition function, as the corresponding matrix model was mapped to the topological string on $\textbf{F}_0$.} It would be interesting to understand if such constant map contributions can be interpreted directly in AdS$_4 \times S^7/\mathbf{Z}_k$ as trivial embeddings of the membrane. 

The full Airy function of the ABJM sphere partition function can thus be understood as coming from the fact that eleven-dimensional supergravity only contributes two terms, the two-derivative action contributing at order $\mu^3$ and its eight-derivative correction contributing at order $\mu$, which have to be inverse Laplace transformed as in \eqref{Eq: inverse laplace of free energy}.\footnote{Note that the vanishing of $\mu^2$ is also vital. Again, this is a direct consequence of scaling arguments with $\ell_p$. Such terms can arise upon introducing localised YM degrees of freedom in the bulk, as in \cite{Horava:1996ma} for example.} It is to be expected that the truncation of subsequent $\mathcal O(1/\mu)$ is a consequence of supersymmetry, and it would be interesting to explain why such eleven-dimensional contributions can indeed not contribute non-trivially to this particular supersymmetric observable.

The identification of the bulk background coupling $A_3$ with the chemical potential $\mu$ not only simplifies the perturbative sector in M-theory, but also the non-perturbative sector. The leading instanton sector functions as a good example in this context. The field theory prediction of its partition function is that of a ratio of Airy functions
\begin{equation}\label{Eq: leading instanton}
	Z_{\text{non-pert.}} = \frac{1}{\sin^2 \frac{2\pi}{k}} \frac{\text{Ai}[\mathcal C^{-1/3}_k (N - \mathcal B_k + 4/k)]}{\text{Ai}[\mathcal C^{-1/3}_k (N - \mathcal B_k )]} + \ldots\,.
\end{equation}
Due to its exponential suppression at large $N$, which goes like $\rme^{-\sqrt{N/k}}$, it was long expected that these instanton contributions are holographically described by membrane instantons wrapping $S^3/\mathbf{Z}_k \subset S^7/\mathbf{Z}_k$ \cite{Drukker:2010nc,Drukker:2011zy}, whose worldvolume metric and on-shell action we provided in \eqref{Eq: metric and on shell action of membrane in AdS4}. And in fact recently this has been further solidified at the quantum level through a one-loop quantisation of such membrane instantons \cite{Beccaria:2023ujc}, following the quantisation of the corresponding worldsheet instantons \cite{Gautason:2023igo}. We note that in this case the chirality constraint in \eqref{Eq: chirality constraint} is solved at a single fixed point, the centre of AdS \cite{Gautason:2025per}. Both charges of the membrane are supersymmetric at this point, and we remember that there are two supersymmetric embeddings inside $S^7/\mathbf{Z}_k$ whose worldvolume metric and on-shell action was already given in \eqref{Eq: metric and on shell action of membrane in AdS4}. Thus there is a degeneracy factor of $4$ multiplying the membrane partition function.

Diagonalising the quadratic membrane action in \eqref{Eq: general quadratic action M2 brane} we find that 
\begin{equation}
\begin{aligned}
	S_{\text{M2}}^{(2)} = \int \rmd^3 x \sqrt{{{g}}} \,\frac12 \bigg( & \sum_{i=1}^{2} D^\alpha \bar\phi_i D_\alpha \phi_i - \frac{3}{4 L^2} \bar \phi_i \phi_i + \rmi\bar \vartheta_i \slashed{D} \vartheta_i - \frac{3 q_{\text{M2}}}{4 L} \bar \vartheta_i \vartheta_i  \\
	&\sum_{i=3}^{4} \nabla^\alpha \bar\phi_i \nabla_\alpha \phi_i 
	+ \rmi\bar \vartheta_i \slashed{\nabla} \vartheta_i + \frac{3q_{\text{M2}}}{4 L} \bar \vartheta_i \vartheta_i\bigg)\,,	
\end{aligned}
\end{equation}
where 
\begin{equation}
	D = \nabla -\rmi q \mathcal A\,,\quad \mathcal A = \frac12 \rmd \varphi - \cos\theta \rmd \phi\,.
\end{equation}
The one-loop determinant of the worldvolume fields evaluates upon $\zeta$-function regularisation to $(4 \sin^2 2\pi/k)^{-1}$. Taking into account the degeneracy factor $4$ the membrane instanton partition function equals
\begin{equation}
	Z_{\text{M2}} = \rme^{-4\mu/k}\frac{1}{\sin^2 \frac{2\pi}{k}}\,,
\end{equation}
correctly reproducing the instanton contribution to the field theory grand canonical potential. Upon inverse Laplace transform, this partition function equates to the ratio of Airy functions in \eqref{Eq: leading instanton}. Explicitly computing the two-loop contribution to the instanton partition function would be an interesting additional non-trivial check for the need of ensemble change.\footnote{A similar two-loop quantisation of a membrane worldvolume partition function, embedded in AdS$_7$ instead, was recently performed in \cite{Beccaria:2025ahf}, where it was also shown to vanish.}

Note that the shift difference in the arguments of the two Airy functions in \eqref{Eq: leading instanton} can be directly explained through the classical on-shell action of the membrane instanton which shifts the argument in the inverse Laplace transform in \eqref{Eq: inverse laplace of free energy}. This same feature also explains the shifts in the partition function of membrane defect operators as well, as we will quickly review in the discussion section.
\subsection{Locally asymptotically AdS$_4$}
As mentioned, utilising four-dimensional minimal gauged $\mathcal N=2$ supergravity allows one to study far more general geometries than empty AdS$_4$. All these geometries still have asymptotically locally AdS boundary conditions. However, their bulk fillings are more involved, and both the internal geometry and the four-form flux get deformed due to the non-trivial four-dimensional gauge field $A$ as shown in \eqref{Eq: 11d metric uplift}. Due to the non-trivial nature of these eleven-dimensional geometries their solutions and physical observables are often studied within the four-dimensional framework. For the purposes of quantising supersymmetric membranes and identifying the canonically conjugate variables and ensembles a direct eleven-dimensional approach is needed instead. This holographic comparison and the choice of ensembles becomes particularly interesting in the context of the so-called Airy conjecture. This conjecture states that the supersymmetric partition function of a generic three-dimensional $\mathcal N=2$ Chern-Simons matter theory on a (squashed) $S^3$ takes the form of an Airy function \cite{Marino:2011eh,Bobev:2022eus,Hristov:2024cgj,Bobev:2025ltz} at large $N$:
\begin{equation}
\begin{aligned}
	Z[N,k,\textbf{q}] =& \,\mathcal C_k(\textbf{q})^{-1/3} \rme^{\mathcal A_k(\textbf{q})} \text{Ai}[\mathcal C_k(\textbf{q})^{-1/3}(N-\mathcal B_k(\textbf{q}))] + \mathcal O (\rme^{-\sqrt{N}})\,
\end{aligned}	
\end{equation}
where $\textbf{q}$ denotes all background couplings determining the theory and the observable that is being computed. The associated grand canonical potential of these theories are thus once more cubic polynomials, up to exponentially suppressed terms, where now the coefficients also depend on the background couplings
\begin{equation}
	J(\mu,k,\textbf{q}) = \frac{\mathcal C_k(\textbf{q})}{3}\mu^3 + \mathcal B_k(\textbf{q}) \mu + \mathcal A_k(\textbf{q}) + \mathcal O(\rme^{-\mu}) \,.
\end{equation}
If true, we thus have that in all the eleven-dimensional geometries dual to these 3d $\mathcal N=2$ vacua the supergravity action contributions are once again confined to two single terms: the standard two-derivative action and its eight-derivative corrections. As a first step to proving this point it would be interesting to verify that the eleven-dimensional action indeed reproduces the $\mu^3$ term in the field theory grand canonical potential instead of the leading $N^{3/2}$ contribution to its free energy, as we have shown above for the round sphere partition function of the ABJM theory.

The Airy conjectures have been generalised to also predict the structure of the superconformal and topologically twisted indices of 3d $\mathcal N=2$ Chern-Simons matter theories, which were argued in the large $N$ limit to take the form of products of Airy functions \cite{Hristov:2024cgj}, again modulo large $N$ instantons. Understanding these structures directly from a bulk analysis, which now contains multiple flux sectors, fixed nuts and bolts loci, can simplify these answers considerably. A first step towards this goal will once again be to directly evaluate the eleven-dimensional supergravity action, and determine the canonically conjugate variables.

The instanton sectors of generic 3d $\mathcal N=2$ CS matter theories have remained largely unexplored, largely due to technical obstructions in analysing the matrix models coming from localising their partition functions. It would be interesting to see if this sector can also be written in a universal manner. As a first step towards studying these non-perturbative sectors more generally we have taken a direct bulk approach and computed the one-loop partition function of the leading worldsheet instanton in the generic asymptotically locally AdS$_4$ geometries presented in \eqref{Eq: 11d metric uplift} \cite{Gautason:2025per}. The on-shell actions and worldvolume metrics of these membrane instantons are universal as described around \eqref{Eq: metric and on shell action of membrane in AdS4}. Its spectrum depends explicitly on the four dimensional gauge field through the following scalar quantity
\begin{equation}\label{Eq: scalar contraction flux}
	f = \frac12 |F + q_{\text{M2}} \star_{4} F| \,,\quad |T| = \frac12 \sqrt{T^{\mu\nu}T_{\mu\nu}}\,,
\end{equation}
in terms of which the masses and charges of the fields are given in Table \ref{Tab: spectrum on membrane}, where the charges $q_1$ and $q_2$ are associated to two independent background gauge fields that enter the covariant derivative
\begin{equation}
	D = \nabla - \rmi q_1 \mathcal A_1 - \rmi q_2 \mathcal A_2\,, \qquad 
		\mathcal A_1 = \frac12 (\rmd \varphi - \cos\theta \rmd \phi)\,,\quad \mathcal A_2 = - \frac12 \cos \theta \rmd \phi \,.
\end{equation}
\begin{table}
\centering
\begin{tabular}{ccccc}
Field & d.o.f. & $q_1$ & $q_2$ & $ML$ \\
\hline\hline
Scalars & 4 & $f$ & 0 & $i f/2$ \\
        & 4 & 1 & 1 & $i\sqrt{3}/2$ \\
\hline
Fermions & 4 & 0 & 0 & $3q_{\text{M2}}/4$ \\
         & 2 & $1+f$ & 1 & $-q_{\text{M2}}(3/4 - f/2)$ \\
         & 2 & $1-f$ & 1 & $-q_{\text{M2}}(3/4 + f/2)$ \\
\hline
\end{tabular}
\caption{Spectrum of worldvolume scalars and fermions on membrane instanton.}\label{Tab: spectrum on membrane}
\end{table}
Once again computing the associated one-loop determinants, regularised with $\zeta$-functions, taking into account the two supersymmetric membranes and their fixed points, we can universally write down the one-loop partition function as \cite{Gautason:2025per}
\begin{equation}\label{Eq: general 1-loop answer}
	Z_{\text{M2}} = 2\sum_{\substack{\text{fixed}\\\text{points}}}\frac{s(\tfrac{2}{k})^{2k}s(x_+)^{-k}s(x_-)^{-k}}{t(x_+)t(x_-)}\rme^{-4 \mu/k}\,, \quad \text{with} \quad x_{\pm} = \frac{2}{k}(1\pm f)\,.
\end{equation}\vspace{-0.4cm}

The functions $s(z)$ and $t(z)$ are defined through a $\zeta$-function regularisation of the following infinite products
\begin{equation}
\begin{aligned}
	&\log s(z) =  \sum\limits_{n=1}^{\infty} n \log\left( \frac{n+z}{n-z} \right) =   {\frac{\rmi  \text{Li}_2(\rme^{2\pi\rmi z})}{2\pi}- \frac{\rmi \pi}{12}  - z\log\left( 1-\rme^{2\pi\rmi z} \right)+\frac{\rmi\pi z^2}{2}}\,,\\
	&t(z) = \prod\limits_{n=1}^{\infty} \frac{k^2}{4}(n^2-z^2) = \frac{4}{k}\frac{\sin\pi z}{z}\,.
\end{aligned}
\end{equation}
\begin{table}\centering
\begin{tabular}{cc|cc}
  Background & Boundary & $f$ & fixed points \\
  \hline\hline
  AdS$_4$ Instanton \cite{Martelli:2011fu} & $S^3_b$ & $\frac{b^2-1}{b^2+1}$ & 1 \\
  Supersymmetric Thermal AdS & $S^2 \times S^1$ & 0 & 0 \\
  AdS-Kerr-Newman Black Hole \cite{Caldarelli:1998hg} & $S^2 \times S^1$ & $\frac{\omega-1}{\omega+1}$ & 2 \\
  Euclidean dyonic $\Sigma_{\mathfrak g>1}$ black hole \cite{Romans:1991nq} & $\Sigma_{\mathfrak g} \times S^1$ & $1$ & $\Sigma_{\mathfrak g}$
\end{tabular}
\caption{Table of specific supersymmetric geometries within 4d minimal gauged supergravity.}\label{Tab: 4d sols}
\end{table}
In Table \ref{Tab: 4d sols} we provide a list of solutions for which the instanton partition function in \eqref{Eq: general 1-loop answer} can be readily evaluated, with below a short description for each example.
\begin{enumerate}\itemsep0em 
	\item \textit{The AdS$_4$ instanton geometry}, dual to the squashed sphere partition function, has a single fixed point at the centre of AdS, just as the empty AdS case. The difference is that only a single chirality of the spinor survives, and subsequently only a single membrane of fixed charge is supersymmetric, leaving a degeneracy factor of $2$ instead of $4$.
	\item \textit{Supersymmetric thermal AdS} does not have any fixed points solving \eqref{Eq: chirality constraint}, and thus there are no membrane instanton contributions.\footnote{Note that also the supergravity action vanishes in this geometry.} Instead the partition function of the dual field theory can be written as a sum over M5-brane giant gravitons \cite{Arai:2020uwd}, which in the bulk was verified at one-loop \cite{Beccaria:2023cuo} for the first giant. Note that in this case no change of ensemble is needed as the boundary observable is computed at fixed $N$, and the M5-brane is quantised at fixed $A_6$. It would be interesting to understand the grand canonical ensemble of the boundary observable in terms of M2-brane dual giant gravitons in the bulk.
	\item The \textit{AdS-Kerr-Newman black hole} has two fixed points, at the poles of the horizon, where $f$ takes identical values, and a single charge of the membrane is supersymmetric, subsequently \eqref{Eq: general 1-loop answer} gets an overall degeneracy factor of $4$.
	\item \textit{The Euclidean dyonic black hole with a Riemann surface $\Sigma_{\mathfrak g>1}$ horizon} is dual to the topologically twisted index. The instanton partition function matches that of the AdS$_4$ instanton background and the Kerr-Newman black hole in the limits $b\rightarrow \infty$ and $\omega \rightarrow 0$, modulo a factor $(1-\mathfrak g)$, which is consistent with \cite{Benini:2016hjo,Choi:2019dfu,Bobev:2023lkx,Bobev:2024mqw}. The additional genus dependent factor arises from the fact that the chirality constraint is solved on the entire Riemann surface horizon, which subsequently needs to be integrated over.
\end{enumerate}
\section{Membranes in AdS$_7 \times S^4$}
Finally, let us review some interesting results in the context of quantum membranes in AdS$_7$ geometries. We will take the internal geometry to simply be $S^4$ such that
\begin{equation}\label{Eq: AdS7 background}
\begin{aligned}
	\rmd s_{11}^2 =&\, L^2\left(\rmd s^2_{\text{AdS}_7} + \frac14 \rmd s_{S^4}^2\right)\,,
\end{aligned}
\end{equation}
although the discussion does generalise to other backgrounds. The background carries a non-trivial three-form gauge potential on the four-sphere which is to be quantised
\begin{equation}
	\rmd A_3 = 3\sqrt{2} L^3  \text{vol}_{S^4} \,,\qquad N = \frac{1}{(2\pi\ell_p)^3} \int_{S^4} \rmd A_3 = \frac{\sqrt{2}}{\pi} \frac{L^3}{\ell_p^3}\,.
\end{equation}
Notably we have that the fixed $A_3$ ensemble, in which we quantise the membrane, agrees with the fixed $N$ ensemble in the dual field theory, the six-dimensional $(2,0)$ theory with gauge group $\text{SU}(N)$. This theory does not contain any coupling constants and is particularly challenging to probe with conventional quantum field theory techniques. It has been proposed that its superconformal index can by computed as the strong coupling limit of the sphere partition function of five-dimensional SYM at radius $r_{S^5}$ \cite{Kim:2012ava,Kim:2012qf,Kim:2012tr,Kim:2013nva}. In this proposal the Yang-Mills gauge coupling is to be identified with the radius of an emergent sixth dimension $4\pi^2/g_{YM}^2 = 1/r_{S^1}$, whose dimensionless periodicity is denoted as $2\pi/\beta = r_{S^5}/r_{S^1}$, and the Yang-Mills instantons are to be mapped to the KK-modes on the emergent circle. In an unrefined $1/4$-BPS limit the partition function was subsequently computed to take the form
\begin{equation}\label{Eq: 2,0 partition function}
	\log Z_{(2,0)}^{S^5\times S^1_{\beta}}[N,\beta] = \frac{\mathcal C_\beta}{3} N^3 + \mathcal B_\beta N + \mathcal A_\beta - \sum_{n,m}^{\infty} \log (1-q^{1+N+m+n})\,,\quad q= \rme^{-\beta}\,,
\end{equation}
where the coefficients in this case equal
\begin{equation}\label{Eq: 2,0 coefficients}
	\mathcal C_\beta = \frac{\beta}{2}\,,\quad \mathcal B_\beta = - \frac{\beta}{8}\,,\quad \mathcal A_\beta = \sum_{n,m}^{\infty} \log (1-q^{1+m+n})\,.
\end{equation}
The relation between the partition function and its associated index goes through a factor of the supersymmetric Casimir energy
\begin{equation}
	Z_{(2,0)}^{S^5\times S^1_{\beta}}[N,\beta] = \rme^{E_c} \mathcal I_{(2,0)}^N[q]\,.
\end{equation}
Evidence that the uplifted SYM partition function can indeed be interpreted as an index comes from the fact that the latter has a proper $q$-expansion. Let us quickly review the origin of the polynomial contributions in $N$. First, there is the perturbative contribution which is computed directly at weak coupling in the five-dimensional Yang-Mills theory
\begin{equation}
	\epsilon_0^{\text{pert.}} = \beta\frac{c_2 |G|}{6} = \beta \frac{N(N^2 - 1)}{6}\,.
\end{equation}
Second, there is a linear term in $N$ which is sensitive to how the Yang-Mills instantons are resummed and expanded at large $\beta$. In fact, as discussed in \cite{Kim:2012ava} there is an ambiguity in doing so. This ambiguity is related to the fact that the partition function itself is ambiguous due to the dimensionfull Yang-Mills coupling, allowing the addition of local terms in the action such as
\begin{equation}
	\frac{\alpha_n}{g_{YM}^{5-2n}} \int_{S^5} \rmd^5 x \sqrt{g}R^n\,,
\end{equation}
where $\alpha_n$ are unconstrained dimensionless constants. In \cite{Kim:2012ava} only $\alpha_2$ is taken non-trivial and linear in $N$ in such a way that the instanton partition function rearranges itself into a dedekind-$\eta$ function. Upon a modular transformation in $\beta \sim g^2_{YM}/r_{S^5}$ the entire partition function can be expanded at large $\beta$ and large $N$ to find an additional contribution of the form
\begin{equation}
	\epsilon_0^{\text{non-pert.}} = \beta \frac{N}{24}\,,
\end{equation}
such that the coefficients $(\mathcal C_\beta,\mathcal B_\beta)$ are as in \eqref{Eq: 2,0 coefficients}, and $\mathcal I_{(2,0)}^N[q]$ indeed has a correct $q$-expansion.

With these caveats in mind we now move to the holographically dual geometry and study to what extent quantum membranes can reproduce this observable. First, since the field theory is on $S^5 \times S^1_{\beta}$ one must use the same boundary in AdS$_7$, which we consequently parametrize as
\begin{equation}
	\rmd s_{\text{AdS}_7}^2 = \rmd \rho^2 + \sinh^2 \rho \,\rmd s_{S^5}^2 + \cosh^2 \rho \,\rmd \tau^2\,.
\end{equation}
The internal geometry does not allow for any stable membrane instantons. Instead, we will study membrane giant gravitons wrapping $S^2\subset S^4$ and $S^1_\beta$, whose supersymmetric embeddings were analysed in \cite{Mikhailov:2000ya}. Note that since $\tau$ is compactified we must impose a twist to globally preserve the background Killing spinors. This twist can be performed with any of the Cartan isometries in the background, but to simplify the discussion we will only impose a twist with the Cartan inside $S^4$ normal to the membrane embedding. In \cite{futurepaperII} we study the generalization of this setup with generic twists preserving a various amount of supersymmetry. The worldvolume metric and on-shell action of the membrane reduces to
\begin{equation}
	\rmd s_{\text{M2}}^2 = L^2(\rmd\tau^2 + \frac14 \rmd s_{S^2}^2)\,,\quad \text{and}\quad S_{\text{M2}}^{S^1_\beta \times S^2} = \beta N
\end{equation}
The quadratic fluctuations arrange themselves into four conformally coupled massless complex scalars and fermions
\begin{equation}\label{Eq: quadratic action giant graviton}
	S = \int \rmd^3 \sigma  \sqrt{g}  \,\sum_{i=1}^{4}\frac12  D \bar \phi_i D \phi_i + \frac12 \bar \phi_i ( -4 \square_{S^2} + 1 ) \phi_i + \bar\vartheta_i (  \slashed{D} + 2  \slashed{\nabla}_{S^2}) \vartheta_i\,,
\end{equation}
where $D$ is along $S^1_\beta$ and contains a flat connection due to the twist
\begin{equation}
	D  = \partial_\tau + q \mathcal A_\tau\,, \quad \text{with charges}\quad q[\phi] = (2,0,0,0)\,,\quad q[\vartheta] = (-1,1,1,1)\,.
\end{equation}
The one-loop partition function of this membrane can be computed by expanding in KK-modes on the circle $S^1_\beta$, which was done in \cite{Beccaria:2023sph} to uplift the worldsheet instanton partition function computed in \cite{Gautason:2023igo}. Alternatively, one can expand in KK-modes on the two-sphere to find that the system reduces to a one-dimensional quantum mechanics. Due to the underlying supersymmetry on the brane, which reduces to an explicit one-dimensional $\mathcal N=2$ algebra, there is a large amount of cancellations of modes whose Witten index trivializes, and one is left with a single complex scalar \cite{futurepaperII}
\begin{equation}
	 S^{(1d)} = \int \rmd \tau\,\frac12 \left(\partial \bar\phi_0 \partial \phi_0 + \bar\phi_0 \phi_0 \right)\,,
\end{equation}
whose partition function is readily computed to equal
\begin{equation}\label{Eq: one-loop partition function}
	Z_{\phi_0} = -\frac{1}{4\sinh^2 \beta/2}\quad \Rightarrow \quad  Z_{\text{M2}} = -\frac{\rme^{-\beta N}}{4\sinh^2 \beta/2}  \,,
\end{equation}
correctly reproducing the result in \cite{Beccaria:2023sph}. This partition function in in complete agreement with the leading instanton contribution in the field theory partition function \eqref{Eq: 2,0 partition function}. Furthermore, due to the exact agreement we conclude that all higher loop corrections on the worldvolume must vanish, very much similar to the membrane instantons in the AdS$_4$ geometries discussed in the previous section. Under the assumption that the supersymmetric cancellation between bosonic and fermionic modes indeed holds when higher order membrane corrections are introduced we are looking at a quantum particle which can be treated in a similar manner as in \cite{Gopakumar:1998ii,Gopakumar:1998jq,Dedushenko:2014nya}, as long as $r_{S^1_\beta}/L$ is large enough and the particle approximation is valid. The partition function of a multi-wound membrane in that case reduces to
\begin{equation}\label{Eq: multi wound giant}
	Z_{\text{M2}}^{(p)} = -\frac{ \rme^{-p\beta N}}{4p\sinh^2 p\beta/2} \,,
\end{equation}
where the rescaling $\beta \rightarrow p \beta$ is due to the multi-winding of the particle, and the additional $1/p$ factor accounts for the cyclic symmetry of the $p$-fold cover of the circle. With the assumption that this free gas approach is valid, and thus no bound states are formed, the membrane partition function correctly reproduces the non-perturbative sector in the field theory partition function \eqref{Eq: 2,0 partition function}
\begin{equation}
	\sum_{p=1}^{\infty} \frac{\rme^{-p\beta N}}{4p\sinh^2 p\beta/2} = \sum_{n,m}^{\infty} \log (1-q^{1+N+m+n})\,.
\end{equation}
So far we have focussed on semi-classically quantising finite sized membranes, but what remains elusive is the dominant supergravity contribution for $L/\ell_p\gg 1$. Let us first start with the constant term $(L/\ell_p)^0$. This is expected to arise from a one-loop correction in M-theory, in analogy with \cite{Gopakumar:1998ii,Gopakumar:1998jq}. In these references the authors studied topological string theory on CY$_3$ manifolds uplifted on a circle to M-theory and these constant contributions were dubbed constant maps, referring to their worldsheet description with trivial embeddings in the CY manifold. It is far from obvious if a similar interpretation is valid here, but with our reduction of the worldvolume theory down to quantum mechanics on the circle we speculate that the partition function of such constant maps take the form as in \eqref{Eq: multi wound giant} albeit without the exponential contribution due to the vanishing classical action for such trivial embeddings. And indeed, this matches the field theory constant map $\mathcal A$ in \eqref{Eq: 2,0 coefficients} contribution, modulo a minus sign.\footnote{Note that a very similar structure can be found in the constant map $\mathcal A_k$ of the ABJM theory in \eqref{Eq: Grand potential}, which in fact can be rewritten as $\mathcal A_k = \sum\limits_{p=1}^{\infty} \frac{k}{4p \sin^2 \frac{\pi k p}{2}} + \frac{\cot \frac{\pi k p}{2}}{2\pi p^2} - \frac{1}{2p \sin^2 \frac{2\pi p}{k}}$, where the latter term is indeed reminiscent of the setup in AdS$_7$ up to a factor two since in AdS$_4$ both the brane and anti-brane are supersymmetric. In the ABJM context the constant map interpretation is more direct as the partition function has a dual description in terms of topological string theory on $\mathbf{F}_0$ \cite{Marino:2009jd}.}

In analogy with the string partition function, one may ask whether the supergravity contributions can also arise from a semi-classical quantisation of degenerate membranes. The difficulty is that this degenerate saddle does not admit a semi-classical treatment, as the membrane instantons. Recent work in the context of topological string theory revisited this issue, and the Dedushenko-Witten Schwinger proper-time approach to membrane instantons \cite{Dedushenko:2014nya}, arguing that the contributions of the degenerate membrane are captured by the zero-pole in the Schwinger integral \cite{Hattab:2023moj,Hattab:2024thi,Hattab:2024ssg}. The proposal to regulate this divergence is by analytically continuing the Schwinger integral in the complex plane and evaluating it as a contour including the positive poles and the pole at the origin. Applied to our holographic setup we effectively resum the entire tower of membrane instantons as a contour integral 
\begin{equation}\label{Eq: complex integral for instantons}
	\sum_{p=1}^{\infty} Z_{\text{M2}}^{(p)} = - \oint_{\mathcal C} \frac{\rmd z}{z}\, K(z)\, \frac{\rme^{-z \beta N}}{4 \sinh^2 z \beta/2}\,,
\end{equation}
where the kernel $K(z)$ is of Mittag-Leffler type, having simple poles at all positive integers
\begin{equation}
	\lim_{z\rightarrow n} K(z) = \frac{1}{2\pi\rmi(z-n)} + \text{reg}_n(z)\,,\qquad n\in \mathbf{N}\,,
\end{equation}
and $\text{reg}_n(z)$ is an entire function. To determine the precise form of this function would require a microscopic understanding of the membrane partition function. Instead of attempting to determine the precise form of this kernel we will set the regular part to zero and use its singular structure to compute the perturbative supergravity data from the zero-pole in the above integral
\begin{equation}
	2\pi \rmi \,\text{Res}_{0}\, K(z)\frac{\rme^{-z \beta N}}{4 z \sinh^2 z \beta/2} = \frac{\beta N(N^2-1)}{6} + \frac{\beta N}{12}\,.
\end{equation}
Interestingly, this answer correctly reproduces the $N^3$ contribution to the $(2,0)$ partition function but differs at subleading order by a term $\beta N/24$. Returning to the derivation of the field theory answer, uplifting the five-dimensional partition function to the $(2,0)$ index, we note that in fact the partition function contains an ambiguity that was fixed by hand in such a way that the Yang-Mills instanton sector contributed an additional term $\beta N/24$ at large $\beta$ and $N$. In the context of the zero-pole it is suggestive that this ambiguity is related to the choice of the entire function $\text{reg}_n(z)$ in the contour prescription above, and it would be very interesting to find a direct UV analysis uniquely determining the kernel.
\section{Final Remarks}
We discussed the quantisation of supersymmetric membranes in holographic geometries, with our main focus going to AdS$_4$ and AdS$_7$ backgrounds. Through holographic dualities we explicitly tested the validity of the semi-classical quantisation of membrane instanton partition functions, and the validity of expanding M-theory partition functions into membrane saddles. These studies highlighted the importance of a choice of ensembles in holographic dualities, controlled by the target space fluxes that are held fixed. Since our interest lies in quantising M2-branes we choose to keep $A_3$ fixed in all backgrounds we studied. In the context of AdS$_4$ holography this meant that the dual (ABJM) field theory is to be studied in a grand canonical ensemble when compared to bulk observables such as membrane instantons and the eleven-dimensional supergravity action, which are all computed at fixed $A_3$. We have tested this explicitly in empty AdS$_4 \times S^7/\mathbf{Z}_k$ and found that both bulk observables equal contributions to the grand canonical potential of the ABJM theory on $S^3$, thus prompting us to conjecture the equality
\begin{equation}
	\mathcal Z_{\text{M2}}[g_{MN},A_3] = J(\mu,k)\,.
\end{equation}
From the boundary field theory it is known that the membrane instantons, uplifted from worldsheet instantons, are one-loop exact in the grand canonical ensemble, motivating an explicit two-loop quantisation of the bulk membrane as a consistency check.

In the context of the Airy conjecture this subsequently lead us to predict that in AdS$_4 \times M_7$ geometries, dual to three-dimensional $\mathcal N=2$ CS matter theories on a squashed three-sphere, the eleven-dimensional supergravity theory only contains two non-trivial contributions to the M-theory partition function: the leading two-derivative action, and its eight-derivative correction.

In addition we discussed the partition function of the leading instanton contribution to a range of ABJM observables, e.g. its squashed sphere partition function, the superconformal index, and topologically twisted index. Semi-classically quantising the associated membrane instantons we found that their partition function takes a universal form with equal exponential suppression and a one-loop contribution that decomposes into (double) sine functions, as per usual for three-dimensional chiral multiplets.

In the context of AdS$_7$ holography we focussed on the 1/2-BPS membrane instanton contributions to the dual $S^5 \times S^1_\beta$ partition function of the six-dimensional $\text{SU}(N)$ $(2,0)$ theory. We found that due to worldvolume supersymmetry there is a large amount of cancellations between fermions and bosons, and that the fluctuations effectively reduce to a quantum mechanical system on $S^1_\beta$. Under the assumption that additional supersymmetric corrections do not spoil this cancellation we argued that the higher order instanton sectors are given simply by multi-wound membranes, up to possible bound states, whose partition function reduces to \eqref{Eq: multi wound giant}. A non-trivial consistency check of this calculation is that the final answer precisely matches the logarithm of the non-perturbative contribution to the boundary partition function.

By resumming the entire tower of membrane instantons into Matsubara modes, written as a contour integral, we discussed the so-called emergence proposal stating that the supergravity contributions to the M-theory partition funcion can be directly extracted from the zero-pole in the contour \cite{Hattab:2024ssg}. We found that indeed this pole correctly reproduces the leading $N^3$ supergravity contribution. At subleading $N^1$ order, corresponding to the eight derivative corrections, there is a mismatch by a term $\beta N/24$, and we discussed its relation to a local counter term ambiguity in the boundary theory \cite{Kim:2012ava,Kim:2013nva}.  

Finally, let us mention that we have not discussed more general membrane observables, such as those associated to defects. In the context of ABJM holography the partition of such defects is one-loop exact in the grand canonical ensemble, but a ratio of Airy functions in the canonical ensemble \cite{Klemm:2012ii}. As we argued, the bulk membrane is to be quantised at fixed $A_3$, and thus compared to the grand canonical ensemble in the field theory, and subsequently we predict that its associated membrane partition function is perturbatively one-loop exact. It would be interesting to explicitly show the two-loop correction to vanish. Such perturbative one-loop exactness in fact also holds in the AdS$_7$ geometry containing defect membranes wrapping an AdS$_3$ \cite{Kim:2012ava,Gautason:2021vfc,Beccaria:2023sph,Beccaria:2025ahf}. 

Let us finish with the observation that all observables we have discussed, in the ensemble of fixed $A_3$, take an interestingly similar form in both the ABJM theory and the six-dimensional $(2,0)$ theory, as we have highlighted in table \ref{Tab: list of observables in AdS4 and AdS7}. This fact might not be that surprising once we understand that all of them arise from quantised membranes in two geometries, i.e. AdS$_4 \times S^7/\mathbf{Z}_k$ and AdS$_{7,\beta} \times S^4$, closely connected through an analytic continuation.

\begin{table}[h]\centering{\renewcommand{\arraystretch}{1.4}
\begin{tabular}{c|c|cc}
    & AdS$_4 \times S^7/\mathbf{Z}_k$ \cite{Fuji:2011km,Hatsuda:2012dt,Klemm:2012ii}  & AdS$_{7,\beta} \times S^4$ \cite{Kim:2012ava}  &\\\hline\hline
$\mathcal Z_{\text{M2}}$ & $\frac{\mathcal C_k}{3} \mu^3 + \mathcal B_k \mu + \mathcal A_k + \mathcal O(\rme^{-\mu})$  & $\frac{\mathcal C_\beta}{3} N^3 + \mathcal B_\beta N + \mathcal A_\beta + \mathcal O(\rme^{-N})$   &                                     \\
$\mathcal C$ & $\frac{2}{\pi^2 k}$  & $\frac{\beta}{2}$   &                                     \\
$\mathcal B$ & $\frac{1}{3k} + \frac{k}{24}$  & $-\frac{\beta}{8}$   &                                     \\
$\mathcal A$ & $\sum\limits_{p=1}^{\infty} \frac{k}{4p \sin^2 \frac{\pi k p}{2}} + \frac{\cot \frac{\pi k p}{2}}{2\pi p^2} - \frac{1}{2p \sin^2 \frac{2\pi p}{k}}$  & $-\sum\limits_{p=1}^{\infty}  \frac{1}{4p \sinh^2 \frac{p\beta}{2}}$   &                                     \\
$\ell$-wound instanton $Z_{\text{M2}}^{(\ell)}$ & $ \frac{\exp (-4\ell\mu/k)}{4 \ell \sin^2 2\pi \ell/k}$ & $\frac{\exp(-\ell N \beta)}{4\ell \sinh^2 \ell\beta/2}$  &  \\
$\ell$-wound defect $\mathcal Z_{\text{M2}}^{(\ell)}$  &  $\frac{\rme (2\ell\mu/k)}{2\sin 2\pi \ell/k}$           &           ${\frac{\exp (\ell\beta N)}{2\sinh \ell\beta/2}} $ & \\
QFT ensemble & grand-canonical & canonical &
\end{tabular}}
\caption{A list of known supersymmetric membrane partition functions, computed in the boundary field theory, both in AdS$_4 \times S^7/\mathbf{Z}_k$ and AdS$_7 \times S^4$ geometries. For the non-perturbatve contributions in AdS$_4$ we have picked out a single tower of membrane instantons wrapping $S^3/\mathbf{Z}_k$, noting that one expects there to be membrane instantons wrapping different internal three cycles, with subsequently different partition functions. Finally, the $\ell$ dependence in the M2-defect in AdS$_7$ has not been derived and can be viewed as a prediction.}\label{Tab: list of observables in AdS4 and AdS7}
\end{table}

%%%%%%%%%%%%%%%%%
\section*{Acknowledgements}
%%%%%%%%%%%%%%%%%
I am thankful Stefan Kuryland, Jieming Lin, Valentina Puletti, Arkady Tseytlin, Zihan Wang, and especially Fridrik Gautason for many interesting discussions and collaborations on quantising strings and branes in holographic backgrounds. JvM is supported by the STFC Consolidated Grant ST/X000575/1. JvM is grateful for the continuous hospitality of the ITF at the KU Leuven.

\bibliography{Corfu_JvM_BIB}

\providecommand{\href}[2]{#2}\begingroup\raggedright\begin{thebibliography}{10}

\bibitem{Fradkin:1984pq}
E.~S. Fradkin and A.~A. Tseytlin, {\it {Effective Field Theory from Quantized Strings}},  {\em Phys. Lett. B} {\bf 158} (1985) 316--322.

\bibitem{Fradkin:1985fq}
E.~S. Fradkin and A.~A. Tseytlin, {\it {Effective Action Approach to Superstring Theory}},  {\em Phys. Lett. B} {\bf 160} (1985) 69--76.

\bibitem{Andreev:1990iv}
O.~D. Andreev, R.~R. Metsaev, and A.~A. Tseytlin, {\it {Covariant calculation of the statistical sum of the two-dimensional sigma model on compact two surfaces.}},  {\em Sov. J. Nucl. Phys.} {\bf 51} (1990) 359--366, [\href{http://arxiv.org/abs/2301.02867}{{\tt arXiv:2301.02867}}].

\bibitem{Tseytlin:1987ww}
A.~A. Tseytlin, {\it {Renormalization of Mobius Infinities and Partition Function Representation for String Theory Effective Action}},  {\em Phys. Lett. B} {\bf 202} (1988) 81--88.

\bibitem{Tseytlin:1988tv}
A.~A. Tseytlin, {\it {Mobius Infinity Subtraction and Effective Action in $\sigma$ Model Approach to Closed String Theory}},  {\em Phys. Lett. B} {\bf 208} (1988) 221--227.

\bibitem{BenettiGenolini:2019jdz}
P.~Benetti~Genolini, J.~M. Perez Ipi\~na, and J.~Sparks, {\it {Localization of the action in AdS/CFT}},  {\em JHEP} {\bf 10} (2019) 252, [\href{http://arxiv.org/abs/1906.11249}{{\tt arXiv:1906.11249}}].

\bibitem{BenettiGenolini:2023kxp}
P.~Benetti~Genolini, J.~P. Gauntlett, and J.~Sparks, {\it {Equivariant Localization in Supergravity}},  {\em Phys. Rev. Lett.} {\bf 131} (2023), no.~12 121602, [\href{http://arxiv.org/abs/2306.03868}{{\tt arXiv:2306.03868}}].

\bibitem{Genolini:2021urf}
P.~B. Genolini and P.~Richmond, {\it {Supersymmetry of higher-derivative supergravity in AdS4 holography}},  {\em Phys. Rev. D} {\bf 104} (2021), no.~6 L061902, [\href{http://arxiv.org/abs/2107.04590}{{\tt arXiv:2107.04590}}].

\bibitem{Gautason:2025per}
F.~F. Gautason and J.~van Muiden, {\it {Localization of the M2-Brane}},  {\em Phys. Rev. Lett.} {\bf 135} (2025), no.~10 101601, [\href{http://arxiv.org/abs/2503.16597}{{\tt arXiv:2503.16597}}].

\bibitem{Gautason:2023igo}
F.~F. Gautason, V.~G.~M. Puletti, and J.~van Muiden, {\it {Quantized strings and instantons in holography}},  {\em JHEP} {\bf 08} (2023) 218, [\href{http://arxiv.org/abs/2304.12340}{{\tt arXiv:2304.12340}}].

\bibitem{Beccaria:2023ujc}
M.~Beccaria, S.~Giombi, and A.~A. Tseytlin, {\it {Instanton contributions to the ABJM free energy from quantum M2 branes}},  {\em JHEP} {\bf 10} (2023) 029, [\href{http://arxiv.org/abs/2307.14112}{{\tt arXiv:2307.14112}}].

\bibitem{Beccaria:2023sph}
M.~Beccaria, S.~Giombi, and A.~A. Tseytlin, {\it {(2,0) theory on $S^5\times S^1$ and quantum M2 branes}},  {\em Nucl. Phys. B} {\bf 998} (2024) 116400, [\href{http://arxiv.org/abs/2309.10786}{{\tt arXiv:2309.10786}}].

\bibitem{Harvey:1999as}
J.~A. Harvey and G.~W. Moore, {\it {Superpotentials and membrane instantons}},  \href{http://arxiv.org/abs/hep-th/9907026}{{\tt hep-th/9907026}}.

\bibitem{Gautason:2025plx}
F.~F. Gautason and J.~van Muiden, {\it {Ensembles in M-theory and holography}},  {\em JHEP} {\bf 11} (2025) 078, [\href{http://arxiv.org/abs/2505.21633}{{\tt arXiv:2505.21633}}].

\bibitem{Aharony:2008ug}
O.~Aharony, O.~Bergman, D.~L. Jafferis, and J.~Maldacena, {\it {N=6 superconformal Chern-Simons-matter theories, M2-branes and their gravity duals}},  {\em JHEP} {\bf 10} (2008) 091, [\href{http://arxiv.org/abs/0806.1218}{{\tt arXiv:0806.1218}}].

\bibitem{Eberhardt:2020bgq}
L.~Eberhardt, {\it {Partition functions of the tensionless string}},  {\em JHEP} {\bf 03} (2021) 176, [\href{http://arxiv.org/abs/2008.07533}{{\tt arXiv:2008.07533}}].

\bibitem{Aharony:2024fid}
O.~Aharony and E.~Y. Urbach, {\it {Type II string theory on AdS$_3 \times S^3\times T^4$ and symmetric orbifolds}},  {\em Phys. Rev. D} {\bf 110} (2024), no.~4 046028, [\href{http://arxiv.org/abs/2406.14605}{{\tt arXiv:2406.14605}}].

\bibitem{Witten:2026twr}
E.~Witten, {\it {Duality and Axion Wormholes}},  \href{http://arxiv.org/abs/2601.01587}{{\tt arXiv:2601.01587}}.

\bibitem{Hattab:2024ssg}
J.~Hattab and E.~Palti, {\it {Notes on integrating out M2 branes}},  {\em Eur. Phys. J. C} {\bf 85} (2025), no.~1 107, [\href{http://arxiv.org/abs/2410.15809}{{\tt arXiv:2410.15809}}].

\bibitem{Blumenhagen:2024lmo}
R.~Blumenhagen, N.~Cribiori, A.~Gligovic, and A.~Paraskevopoulou, {\it {Reflections on an M-theoretic Emergence Proposal}},  {\em PoS} {\bf CORFU2023} (2024) 238, [\href{http://arxiv.org/abs/2404.05801}{{\tt arXiv:2404.05801}}].

\bibitem{Candelas:1990rm}
P.~Candelas, X.~C. De~La~Ossa, P.~S. Green, and L.~Parkes, {\it {A Pair of Calabi-Yau manifolds as an exactly soluble superconformal theory}},  {\em Nucl. Phys. B} {\bf 359} (1991) 21--74.

\bibitem{Gopakumar:1998ii}
R.~Gopakumar and C.~Vafa, {\it {M theory and topological strings. 1.}},  \href{http://arxiv.org/abs/hep-th/9809187}{{\tt hep-th/9809187}}.

\bibitem{Gopakumar:1998jq}
R.~Gopakumar and C.~Vafa, {\it {M theory and topological strings. 2.}},  \href{http://arxiv.org/abs/hep-th/9812127}{{\tt hep-th/9812127}}.

\bibitem{futurepaper}
F.~F. Gautason and J.~van Muiden, {\it {work in progress}}, .

\bibitem{futurepaperII}
F.~F. Gautason and J.~van Muiden, {\it {work in progress}}, .

\bibitem{Bergshoeff:1987cm}
E.~Bergshoeff, E.~Sezgin, and P.~K. Townsend, {\it {Supermembranes and Eleven-Dimensional Supergravity}},  {\em Phys. Lett. B} {\bf 189} (1987) 75--78.

\bibitem{Gauntlett:2007ma}
J.~P. Gauntlett and O.~Varela, {\it {Consistent Kaluza-Klein reductions for general supersymmetric AdS solutions}},  {\em Phys. Rev. D} {\bf 76} (2007) 126007, [\href{http://arxiv.org/abs/0707.2315}{{\tt arXiv:0707.2315}}].

\bibitem{Drukker:2010nc}
N.~Drukker, M.~Marino, and P.~Putrov, {\it {From weak to strong coupling in ABJM theory}},  {\em Commun. Math. Phys.} {\bf 306} (2011) 511--563, [\href{http://arxiv.org/abs/1007.3837}{{\tt arXiv:1007.3837}}].

\bibitem{Marino:2011eh}
M.~Marino and P.~Putrov, {\it {ABJM theory as a Fermi gas}},  {\em J. Stat. Mech.} {\bf 1203} (2012) P03001, [\href{http://arxiv.org/abs/1110.4066}{{\tt arXiv:1110.4066}}].

\bibitem{Drukker:2011zy}
N.~Drukker, M.~Marino, and P.~Putrov, {\it {Nonperturbative aspects of ABJM theory}},  {\em JHEP} {\bf 11} (2011) 141, [\href{http://arxiv.org/abs/1103.4844}{{\tt arXiv:1103.4844}}].

\bibitem{Hatsuda:2012dt}
Y.~Hatsuda, S.~Moriyama, and K.~Okuyama, {\it {Instanton Effects in ABJM Theory from Fermi Gas Approach}},  {\em JHEP} {\bf 01} (2013) 158, [\href{http://arxiv.org/abs/1211.1251}{{\tt arXiv:1211.1251}}].

\bibitem{Hatsuda:2012hm}
Y.~Hatsuda, S.~Moriyama, and K.~Okuyama, {\it {Exact Results on the ABJM Fermi Gas}},  {\em JHEP} {\bf 10} (2012) 020, [\href{http://arxiv.org/abs/1207.4283}{{\tt arXiv:1207.4283}}].

\bibitem{Hatsuda:2013gj}
Y.~Hatsuda, S.~Moriyama, and K.~Okuyama, {\it {Instanton Bound States in ABJM Theory}},  {\em JHEP} {\bf 05} (2013) 054, [\href{http://arxiv.org/abs/1301.5184}{{\tt arXiv:1301.5184}}].

\bibitem{Fuji:2011km}
H.~Fuji, S.~Hirano, and S.~Moriyama, {\it {Summing Up All Genus Free Energy of ABJM Matrix Model}},  {\em JHEP} {\bf 08} (2011) 001, [\href{http://arxiv.org/abs/1106.4631}{{\tt arXiv:1106.4631}}].

\bibitem{Dedushenko:2014nya}
M.~Dedushenko and E.~Witten, {\it {Some Details On The Gopakumar-Vafa and Ooguri-Vafa Formulas}},  {\em Adv. Theor. Math. Phys.} {\bf 20} (2016) 1--133, [\href{http://arxiv.org/abs/1411.7108}{{\tt arXiv:1411.7108}}].

\bibitem{Horava:1996ma}
P.~Horava and E.~Witten, {\it {Eleven-dimensional supergravity on a manifold with boundary}},  {\em Nucl. Phys. B} {\bf 475} (1996) 94--114, [\href{http://arxiv.org/abs/hep-th/9603142}{{\tt hep-th/9603142}}].

\bibitem{Beccaria:2025ahf}
M.~Beccaria, S.~A. Kurlyand, and A.~A. Tseytlin, {\it {2-loop free energy of M2 brane in AdS$_7 \times$ S$^4$ and surface defect anomaly in (2,0) theory}},  \href{http://arxiv.org/abs/2511.22306}{{\tt arXiv:2511.22306}}.

\bibitem{Bobev:2022eus}
N.~Bobev, J.~Hong, and V.~Reys, {\it {Large N partition functions of the ABJM theory}},  {\em JHEP} {\bf 02} (2023) 020, [\href{http://arxiv.org/abs/2210.09318}{{\tt arXiv:2210.09318}}].

\bibitem{Hristov:2024cgj}
K.~Hristov, {\it {Equivariant localization and gluing rules in 4d $\mathcal{N}=2$ higher derivative supergravity}},  6, 2024.
\newblock \href{http://arxiv.org/abs/2406.18648}{{\tt arXiv:2406.18648}}.

\bibitem{Bobev:2025ltz}
N.~Bobev, P.-J. De~Smet, J.~Hong, V.~Reys, and X.~Zhang, {\it {An Airy Tale at Large $N$}},  \href{http://arxiv.org/abs/2502.04606}{{\tt arXiv:2502.04606}}.

\bibitem{Martelli:2011fu}
D.~Martelli, A.~Passias, and J.~Sparks, {\it {The gravity dual of supersymmetric gauge theories on a squashed three-sphere}},  {\em Nucl. Phys. B} {\bf 864} (2012) 840--868, [\href{http://arxiv.org/abs/1110.6400}{{\tt arXiv:1110.6400}}].

\bibitem{Caldarelli:1998hg}
M.~M. Caldarelli and D.~Klemm, {\it {Supersymmetry of Anti-de Sitter black holes}},  {\em Nucl. Phys. B} {\bf 545} (1999) 434--460, [\href{http://arxiv.org/abs/hep-th/9808097}{{\tt hep-th/9808097}}].

\bibitem{Romans:1991nq}
L.~J. Romans, {\it {Supersymmetric, cold and lukewarm black holes in cosmological Einstein-Maxwell theory}},  {\em Nucl. Phys. B} {\bf 383} (1992) 395--415, [\href{http://arxiv.org/abs/hep-th/9203018}{{\tt hep-th/9203018}}].

\bibitem{Arai:2020uwd}
R.~Arai, S.~Fujiwara, Y.~Imamura, T.~Mori, and D.~Yokoyama, {\it {Finite-$N$ corrections to the M-brane indices}},  {\em JHEP} {\bf 11} (2020) 093, [\href{http://arxiv.org/abs/2007.05213}{{\tt arXiv:2007.05213}}].

\bibitem{Beccaria:2023cuo}
M.~Beccaria and A.~A. Tseytlin, {\it {Large N expansion of superconformal index of k=1 ABJM theory and symiclassical M5 brane partition function}},  {\em Nucl. Phys. B} {\bf 1001} (2024) 116507, [\href{http://arxiv.org/abs/2312.01917}{{\tt arXiv:2312.01917}}].

\bibitem{Benini:2016hjo}
F.~Benini and A.~Zaffaroni, {\it {Supersymmetric partition functions on Riemann surfaces}},  {\em Proc. Symp. Pure Math.} {\bf 96} (2017) 13--46, [\href{http://arxiv.org/abs/1605.06120}{{\tt arXiv:1605.06120}}].

\bibitem{Choi:2019dfu}
S.~Choi and C.~Hwang, {\it {Universal 3d Cardy Block and Black Hole Entropy}},  {\em JHEP} {\bf 03} (2020) 068, [\href{http://arxiv.org/abs/1911.01448}{{\tt arXiv:1911.01448}}].

\bibitem{Bobev:2023lkx}
N.~Bobev, J.~Hong, and V.~Reys, {\it {Large N partition functions of 3d holographic SCFTs}},  {\em JHEP} {\bf 08} (2023) 119, [\href{http://arxiv.org/abs/2304.01734}{{\tt arXiv:2304.01734}}].

\bibitem{Bobev:2024mqw}
N.~Bobev, S.~Choi, J.~Hong, and V.~Reys, {\it {Superconformal indices of 3d $ \mathcal{N} $ = 2 SCFTs and holography}},  {\em JHEP} {\bf 10} (2024) 121, [\href{http://arxiv.org/abs/2407.13177}{{\tt arXiv:2407.13177}}].

\bibitem{Kim:2012ava}
H.-C. Kim and S.~Kim, {\it {M5-branes from gauge theories on the 5-sphere}},  {\em JHEP} {\bf 05} (2013) 144, [\href{http://arxiv.org/abs/1206.6339}{{\tt arXiv:1206.6339}}].

\bibitem{Kim:2012qf}
H.-C. Kim, J.~Kim, and S.~Kim, {\it {Instantons on the 5-sphere and M5-branes}},  \href{http://arxiv.org/abs/1211.0144}{{\tt arXiv:1211.0144}}.

\bibitem{Kim:2012tr}
H.-C. Kim and K.~Lee, {\it {Supersymmetric M5 Brane Theories on R x CP2}},  {\em JHEP} {\bf 07} (2013) 072, [\href{http://arxiv.org/abs/1210.0853}{{\tt arXiv:1210.0853}}].

\bibitem{Kim:2013nva}
H.-C. Kim, S.~Kim, S.-S. Kim, and K.~Lee, {\it {The general M5-brane superconformal index}},  \href{http://arxiv.org/abs/1307.7660}{{\tt arXiv:1307.7660}}.

\bibitem{Mikhailov:2000ya}
A.~Mikhailov, {\it {Giant gravitons from holomorphic surfaces}},  {\em JHEP} {\bf 11} (2000) 027, [\href{http://arxiv.org/abs/hep-th/0010206}{{\tt hep-th/0010206}}].

\bibitem{Marino:2009jd}
M.~Marino and P.~Putrov, {\it {Exact Results in ABJM Theory from Topological Strings}},  {\em JHEP} {\bf 06} (2010) 011, [\href{http://arxiv.org/abs/0912.3074}{{\tt arXiv:0912.3074}}].

\bibitem{Hattab:2023moj}
J.~Hattab and E.~Palti, {\it {On the particle picture of Emergence}},  {\em JHEP} {\bf 03} (2024) 065, [\href{http://arxiv.org/abs/2312.15440}{{\tt arXiv:2312.15440}}].

\bibitem{Hattab:2024thi}
J.~Hattab and E.~Palti, {\it {Emergence in string theory and Fermi gases}},  {\em JHEP} {\bf 07} (2024) 144, [\href{http://arxiv.org/abs/2404.05176}{{\tt arXiv:2404.05176}}].

\bibitem{Klemm:2012ii}
A.~Klemm, M.~Marino, M.~Schiereck, and M.~Soroush, {\it {Aharony\textendash{}Bergman\textendash{}Jafferis\textendash{}Maldacena Wilson Loops in the Fermi Gas Approach}},  {\em Z. Naturforsch. A} {\bf 68} (2013), no.~1-2 178--209, [\href{http://arxiv.org/abs/1207.0611}{{\tt arXiv:1207.0611}}].

\bibitem{Gautason:2021vfc}
F.~F. Gautason and V.~G.~M. Puletti, {\it {Precision holography for 5D Super Yang-Mills}},  {\em JHEP} {\bf 03} (2022) 018, [\href{http://arxiv.org/abs/2111.15493}{{\tt arXiv:2111.15493}}].

\end{thebibliography}\endgroup
\bibliographystyle{JHEP}

\end{document}